\documentclass[twocolumn,secnumarabic,amssymb, amsmath, amsfonts , nofootinbib,tightenlines,showpacs, aps, prl,superscriptaddress]{revtex4}
\usepackage{amsfonts}
\usepackage{amsmath}
\usepackage{amssymb}
\usepackage{graphicx}%
\usepackage{bm}%
\usepackage{graphicx}
\begin{document}
\input{epsf}
\def\Fig#1{Figure \ref{#1}}
\def\Eq#1{Eq.~(\ref{#1})}
\def\Table#1{Table~\ref{#1}}

\title[Equilibrium Electro-osmotic Instability]{Equilibrium Electro-osmotic Instability}

\author{I. Rubinstein}\email{robinst@post.bgu.ac.il.}\affiliation{Blaustein Institutes for Desert
Research, Ben-Gurion University of the Negev, Sede Boqer Campus
84993, Israel}
\author{B. Zaltzman }\email{boris@bgu.ac.il.}\affiliation{Blaustein Institutes for Desert Research, Ben-Gurion
University of the Negev, Sede Boqer Campus 84993, Israel}


\begin{abstract}
Since its prediction fifteen years ago, electro-osmotic instability has been attributed to non-equilibrium electro-osmosis related to the extended space charge which develops at the limiting current in the course of concentration polarization at a charge-selective interface. This attribution had a double basis. Firstly, it has been recognized that equilibrium electro-osmosis cannot yield instability for a perfectly charge-selective solid. Secondly, it has been shown that non-equilibrium electro-osmosis can. First theoretical studies in which electro-osmotic instability was predicted and analyzed employed the assumption of perfect charge-selectivity for the sake of simplicity and so did the subsequent numerical studies of various time-dependent and nonlinear features of electro-osmotic instability. In this letter, we show that relaxing the assumption of perfect charge-selectivity (tantamount to fixing the electrochemical potential in the solid) allows for equilibrium electro-osmotic instability. Moreover, we suggest a simple experimental test for determining the true, either equilibrium or non-equilibrium, origin of electro-osmotic instability.
\end{abstract}

\maketitle Recently, electro-osmotic instability (EOI) in concentration polarization (CP) at a charge-selective solid has attracted a considerable interest of both theoreticians and experimenters, \cite{1}--\cite{8}. Since its prediction 15 years ago, \cite{9}, EOI has been attributed to non-equilibrium electro-osmosis (EO) related to the extended space charge (ESC), \cite{10}--\cite{13}, which develops in the course of CP at the limiting current (LC). This attribution had a double basis. Firstly, it has been recognized that equilibrium EO cannot yield instability for a perfectly perm-(charge-)selective solid, \cite{14}. Secondly, it has been shown that non-equilibrium EO can, \cite{15}. First theoretical studies in which EOI was predicted and analyzed employed the assumption of perfect perm-selectivity for the sake of simplicity \cite{9}, \cite{13}, \cite{15},  and so did the subsequent numerical studies of various time-dependent and nonlinear features of EOI, \cite{2}, \cite{3}, \cite{5}, \cite{16}. In this letter, we show that relaxing the assumption of perfect perm-selectivity (tantamount to fixing the electrochemical potential in the solid) allows for equilibrium EOI. Moreover, we suggest a simple experimental test for determining the true, either equilibrium or non-equilibrium, origin of EOI.

DC ionic current in a binary electrolyte passing through a perm-selective interface (electrode, ion exchange membrane, micro-nano-channel junction) is a basic element of many electrochemical engineering or micro-fluidic processes, such as electrodeposition, electrodialysis, or protein pre-concentration, \cite{17}, \cite{18}. Such current passage is diffusion limited in the sense that it induces a decrease of electrolyte concentration towards the interface, known as the ionic CP. A common expression of CP is a characteristic voltage-current (VC) curve with a segment in which the current nearly saturates at some plateau value, LC, corresponding to nearly vanishing interface concentration. This segment of the VC curve is usually followed by another region of a relatively rapid increase of electric current with voltage –- the so-called over-limiting conductance (OLC) regime. The mechanism of OLC remained unexplained for a long time. Only recently was it shown that in open systems OLC is due to destruction of the diffusion layer (DL) by a micro-scale interface driven vortical flow which spontaneously develops as a result of instability of  CP near the LC and provides an additional ionic transport mechanism yielding OLC, \cite{4}, \cite{7}, \cite{8}, \cite{15}, \cite{16}, \cite{19}, \cite{20}. This flow is driven by the electric force acting upon the space charge of a nanometers-thick interfacial electric double layer (EDL). A slip-like fluid flow induced by this force is known as EO. There are two regimes of EO that correspond to different states of EDL and are controlled by the non-equilibrium voltage drop (overvoltage) across it, \cite{8}: equilibrium EO, \cite{9}, \cite{10} and non-equilibrium EO, or EO of the Second Kind, \cite{2}, \cite{8}, \cite{11}. While both regimes result from the action of a tangential electric field upon the space charge of EDL, the former relates to the charge of equilibrium EDL, whereas the latter relates to ESC of non-equilibrium EDL which develops in the course of CP near LC.

The theory of equilibrium EO at a perm-selective interface was developed by Dukhin and Derjaguin \cite{21}. An essential component of this theory is accounting for polarization of the EDL by the applied tangential electric field, resulting in a lateral pressure drop in the double layer, owing to the lateral variation of Maxwell stress. This yields for equilibrium EO slip velocity, instead of the common Helmholtz-Smoluchowski formula  $u=-\zeta E$, the expression
\begin{equation}
u=\zeta\left({\varphi}_y+\frac{{c}_y}{{c}}\right)+\frac{{c}_y}{{c}}\left(4\ln2-4\ln\left(e^{\zeta/2}+1\right)\right).\label{1}
\end{equation}
Here $\zeta$ is the electric potential drop between the interface and the outer edge of EDL. The peculiarity of (\ref{1}) is that, for an ideally perms-elective cation exchange membrane maintained at a fixed electric potential, the electrochemical potential of counterions in the membrane, $\ln c+\varphi=\textrm{const.}$, is constant, and so it is, in equilibrium conditions, at the outer edge of EDL. In other words, $\partial {c}/\partial y=-c\partial {\varphi}/\partial y$, and for $\zeta\to-\infty$, equation (\ref{1}) yields
\begin{equation}
{u}=-4\ln2 {\varphi}_y\label{2}
\end{equation}
That is, the factor at $-\partial {\varphi}/\partial y$ (EO factor) tends to a maximal upper value upon the increase of $\zeta$ (negative). This stands in contrast with the respective prediction of the Helmholtz-Smoluchowski formula and is a direct consequence of polarization of the EDL at a perm-selective interface. Hydrodynamic stability of the quiescent concentration polarization with a limiting equilibrium EO slip condition (\ref{2}), was studied by Zholkovskij et al., \cite{14}, who concluded that 1D CP was stable. On the other hand, it was subsequently shown that the non-equilibrium slip related to ESC did yield instability, \cite{9}, \cite{13}, \cite{15}. This was the reason why since its prediction in 1999, \cite{9}, till now EOI was attributed to non-equilibrium EO and was so studied, \cite{1}--\cite{6}, \cite{13}, \cite{18}.

It is the purpose of this letter to show that any deviations from constancy of the electrochemical potential of counter-ions at the outer edge of EDL makes EOI possible with equilibrium EO slip, (\ref{1}). Non-constancy of the counter-ionic electrochemical potential may result either from non-ideal perm-selectivity of the interface (non-ideally perm-selective nano-slot or ion-exchange membrane),  addressed in this letter, or from a finite rate of electrode reactions (e.g., in cathodic deposition). This letter is structured as follows.  We begin by formulating a three-layer model for a membrane flanked by two CP-polarized DLs whose stability under equilibrium slip condition (\ref{1}) we analyze. Next, the results of a linear stability analysis in this model are presented, followed by the results of illustrative numerical simulations in the full non-linear model.

\begin{figure}[h]
\includegraphics[width=\columnwidth,keepaspectratio=true]{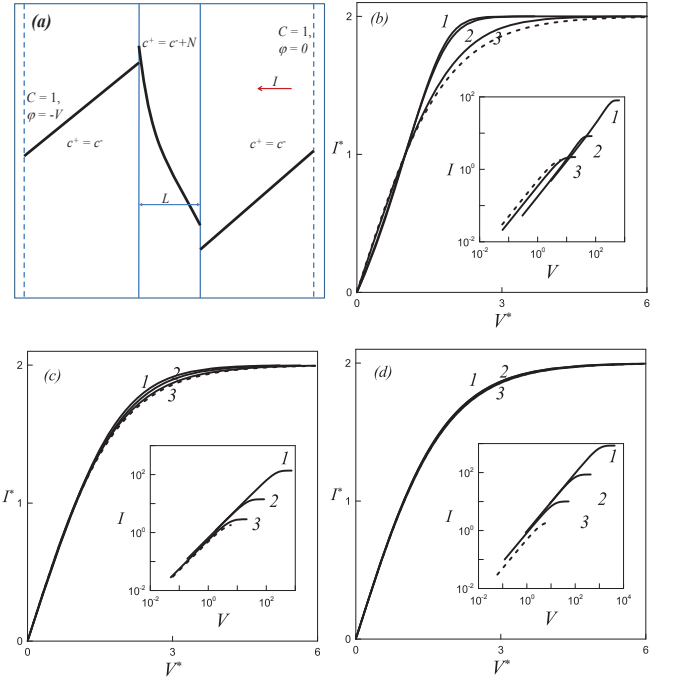}
\caption{(a) Scheme of three-layer setup, bold lines -- schematic plots of salt concentration, $C(x)$; (b)--(d) Scaled VC-dependencies: (a) $L=10$, (b) $L=1$, (c) $L=0.1$; (1) $N=0.1$, (2) $N=1$, (3) $N=10$, dashed line corresponds to perfectly perm-selective solid. Insets: Same plots for unscaled VC-dependencies.}
\end{figure}

 Let us consider a 2D cation-exchange membrane (nano-channel array), $0<x<L$ flanked by two DLs, $-1<x<0$ and $L<x<L+1$, of a univalent electrolyte with concentration $c_0$ maintained at the outer boundary of DLs (see Fig.1a,b). This three-layer system is modeled by the following boundary-value problem non-dimensionlized in a natural manner, \cite{15},
\begin{align}
&\frac{\partial c_\pm}{\partial t}=-\mathbf{\nabla}\cdot{\mathbf{j_\pm}};\label{3}\\
&\mathbf{j}_{\pm}=-c_{\pm}\mathbf{\nabla}\mu_{\pm}-\textrm{Pe} \mathbf{v}c_{\pm}, \ \mu_{\pm}=\ln{c_{\pm}}\pm \varphi;\label{4}\
\end{align}
where $c_\pm$ is concentration of positive and negative ions, $\textrm{Pe}$ is material Peclet number (electric Rayleigh number), \cite{15}, and $\varphi$ is electric potential scaled to the thermal voltage, $kT/e$. Electroneutrality conditions in the enriched, $-1<x<0$, and in the depleted, $L<x<L+1$ DLs and the membrane, $0<x<L$, read, respectively:
\begin{align}
&c_+=c_-=C,\ -1<x<0, \ L<x<L+1;\label{5}\\
&c_+=c_-+N=C+\frac{N}{2},\ 0<x<L.\label{6}\
\end{align}
Here $N$ is dimensionless fixed charge density in the membrane. Let us neglect the fluid flow in the enriched DL and in the membrane: $\mathbf{v}=w\mathbf{i}+u\mathbf{j}\equiv 0,\ -1<x<L$; and determine it in the depleted DL from the Stokes-continuity equations:
\begin{equation}
\mathbf{\nabla}^2 \mathbf{v}-\mathbf{\nabla}p=0,\ \mathbf{\nabla}\cdot\mathbf{v}=0.\label{7}
\end{equation}
At the outer boundaries of DLs we assume vanishing fluid velocity, $\mathbf{v}=0,\ x=L+1 $ and prescribe the dimensionless concentration and the electric potential:
\begin{equation}
C|_{x=-1,L+1}=1,\ \varphi|_{x=-1}=-V,\ \varphi|_{x=L+1}=0.\label{8}
\end{equation}
We complete the formulation by prescribing continuity of electrochemical potentials, $\mu_{\pm}$, and normal ionic fluxes through the membrane--solution interfaces, $x=0,L$ and the slip condition (\ref{1}) at the membrane--depleted DL interface, $x=L$. The main control parameters are  dimensionless voltage $V$, dimensionless width of the membrane $L$,  and $N$ -- dimensionless fixed charge density in the membrane.
\begin{figure}[h]
\hspace{-0.75cm}
\includegraphics[width=4in,keepaspectratio=true]{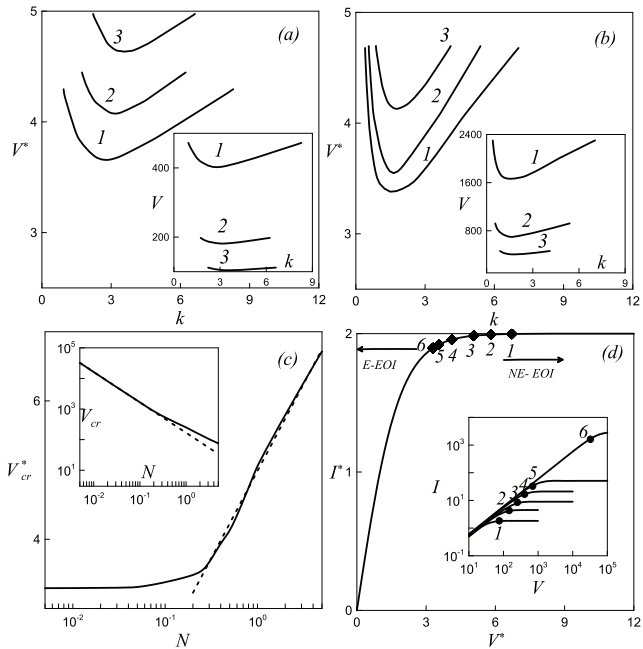}
\vspace{-1.25cm}
\caption{Neutral stability curves in scaled voltage $V^*$--wave number $k$ plane (above the curve -- instability): (a) $L=1$, (b) $L=0.1$; $N=0.1(1)$, $N=1(2)$, $N=10(3)$; Insets: Same for unscaled voltage $V$. (c) Scaled critical voltage, $V^*_{cr}$, versus $N$, $L=0.1$. Dashed line stands for logarithm fit. Inset: Same for $V_{cr}$. Dashed line stands for $1/N$-power fit. (d) Instability threshold on the $V^*-I^*$ curve for $L=0.1$ and $N=5(1)$, $N=2(2)$, $N=1(3)$, $N=0.5(4)$, $N=0.25(5)$, $N=0.005(6)$. Arrows mark the variation of threshold with decreasing $N$ (increasing bulk concentration in dimensional terms) for equilibrium and non-equilibrium EOI.  Inset: Same six points on the $V-I$ curves.}
\end{figure}

Quiescent 1D steady-state solution to the problem (\ref{1}), (\ref{3}--\ref{8}) has been computed analytically in terms of Lambert functions. In Fig.1b--c we present the computed VC dependencies for various $N$ and $L$ (current density $\mathbf{I},\ \mathbf{I}=\mathbf{j}^+-\mathbf{j}^-$; in figures below $I$ is the normal component of $\mathbf {I}$ averaged over the interface). We note that whereas the VC curves computed for different $N$ and $L$ strongly differ due to a decrease of membrane perm-selectivity with the decrease of $N$ or $L$ (insets to Figs.1b--c),  upon suitable scaling, the scaled $I^*-V^*$ curves collapse and converge to the scaled VC dependence for a perfectly-perm-selective solid interface (dashed line, Figs.1b--c). Here $I^*=I/I_0,\ V^*=V/V_0$, where $I_0$ is , e.g., one half of LC, $I_{lim}$, and $V_0$ is the corresponding voltage.

The results of a linear stability analysis of the quiescent 1D steady-state are presented in Fig.2. The region above the neutral-stability curves plotted in Fig.2a,b corresponds to instability. Whereas for unscaled voltage the unstable portion of the $V$-$k$ plane shrinks exponentially with a decrease of membrane perm-selectivity (insets to Fig.2a,b), for scaled voltage the unstable portion of the voltage-wave number phase plane expands with decreasing $N$. The dependence of critical voltage on $N$ is depicted in Fig.2c. In terms of the unscaled voltage the $N^{-1}$-dependence is observed for a weakly selective solid (dashed line in inset to Fig.2c), whereas the scaled $V^*_{cr}$ behaves as $\ln N$ for a strongly perm-selective solid and converges asymptotically to some finite value for $N<<1$. This decrease of the scaled critical voltage with saturation at low $N$ is a particular feature of equilibrium EOI. This point is potentially important for experimental identification of the origin of the OLC flow. To emphasize this, on the scaled VC curve we mark six points corresponding to the onset of instability for a decreasing sequence of $N$ (in dimensional terms, decrease of $N$ is tantamount to increasing bulk solute concentration), Fig.2d, along with the corresponding plot in the unscaled $V-I$-plane (inset to Fig.2d). We note that, whereas for a perfectly perm-selective membrane and its related  non-equilibrium EOI, the increase of bulk electrolyte concentration, resulting in the decrease of the dimensionless Debye length, yields an increase of scaled threshold voltage, \cite{13}, for equilibrium EOI under consideration, the increase of bulk concentration yields a decrease of $N$ accompanied by a decrease of scaled threshold voltage $V^*$, Fig.2d. This feature may serve a tool for experimental identification of the OLC mechanism.

Examining the leading mode of the perturbed solution we note that the EO, $\zeta\varphi_y$, and the diffusio-osmotic, $\zeta c_y/c$, terms in the slip condition (\ref{1}) have opposite signs: whereas the former is driving the equilibrium EOI the latter plays a stabilizing role, Fig.3a. As a result, with saturation of counter-ion electrochemical potential at the membrane-solution interface upon the increase of membrane's perm-selectivity, Fig.3a, and its related convergence of slip velocity to its asymptotic limit (\ref{2}), inset to Fig.3a, the second  diffusio-osmotic term balances the first one in the slip condition (\ref{1}), and the threshold voltage for equilibrium EOI increases unboundedly for $N$ tending to infinity.

In Fig.3b-e we illustrate the OLC resulting from equilibrium EOI. Thus, in Fig.3b we present a numerically computed VC dependence. We note the good agreement between the critical voltage and wave number predicted by the linear stability analysis with those obtained by numerical solution of the nonlinear problem  (\ref{1}), (\ref{3})--(\ref{8}), Fig.3b-e.

\begin{figure}
\includegraphics[width=4in,keepaspectratio=true]{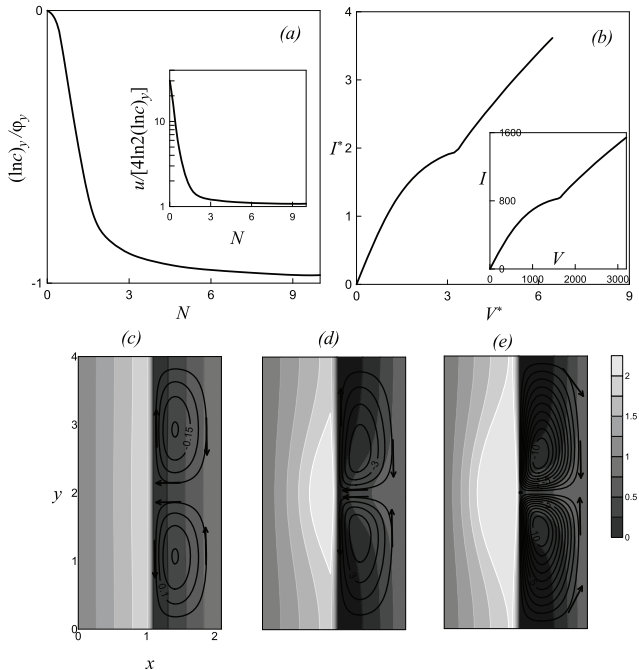}
\vspace{-1.25cm}
\caption{(a) Ratio of tangential variation of the perturbed $\varphi$ and $\ln c$ at the membrane--depleted DL interface versus $N$; $L=0.1$, and $V^*=V^*_{cr},\ k=k_{cr}$. Inset: ratio of slip velocity $u$ and its asymptotic value (\ref{2}) computed from linear stability analysis. (b) Steady-state $V^*-I^*$ dependence; $N=0.1,\ L=0.1$. Inset: Unscaled VC dependence. (c)--(e) Concentration distribution (darker color corresponds to lower concentration), and flow streamlines computed for $N=0.1,\ L=0.1$, (c) $V^*=3.4\sim V^*_{cr}$, (d) $V^*=4$, (e) $V^*=6.3$.}
\end{figure}

In the OLC regime for high currents, $I\ge 1.7 I_{lim}$ ($L=0.1,\ N=0.1$), an additional pair of small vortexes appears near the stagnation point, $x=1.1,\ y=2$, between the descending portions of the EOI vortexes, Fig.3e and Fig.4b,c. Further increase of voltage yields a break-down of vortex symmetry and transition to chaotic oscillations, Fig.4a.

\begin{figure}
\includegraphics[width=3.5in,keepaspectratio=true]{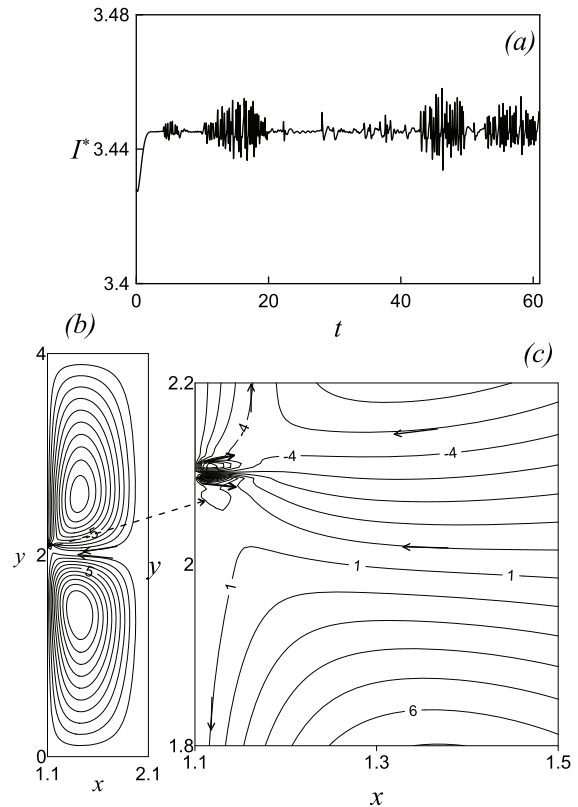}
\vspace{-1.25cm}
\caption{(a) Chaotic oscillations of $I^*$ for $N=0.1,\ L=0.1, V^*=6.5$. (b) Streamlines; asymmetry of vortexes. (c) Streamlines; vicinity of the stagnation point.}
\end{figure}

To conclude, we point out that equilibrium EOI rather than its non-equilibrium counterpart may be accountable for the  reported effects of depleted membrane surface modification by surfactants and polyelectrolyte deposition upon the onset of OLC, \cite{22}.

The work was supported by the USA-Israel Binational Science Foundation (grant 2010199).


\begin{thebibliography}{100}
\bibitem[20]{20} This pertains to nonconfined systems. Geometric confinement changes the mechanism of OLC in CP. In very shallow micro-channels (depth on a micron/submicron scale) preceding the nano-channel (membrane), OLC is associated with the surface conduction effect. As the micro-channel depth is increased to the tens of microns range, another mechanism –- convection due to EO flow along the charged channel walls – eventually overtakes the surface conduction. For a dead-end channel sealed by an impermeable membrane, pressure-driven flow develops to oppose EO flow and results in the formation of two counter-rotating vortices, which carry concentrated solution to the membrane along the sidewalls
and return depleted solution along the channel axis. This circulation constitutes the dominant mechanism of OLC for the intermediate range of confinements. Upon a further reduction of confinement with, transition to the EOI mechanism of OLC occurs.
\end{thebibliography}
\end{document}